\documentclass[12pt]{article} 
\usepackage{graphicx,epsfig}
\begin{document} 

\begin{center}
{\bf \large Rotational friction on small globular proteins: Combined dielectric and 
hydrodynamic effect}

{\bf Arnab Mukherjee and Biman Bagchi \footnote[1]{Email: bbagchi@sscu.iisc.ernet.in}}

{\large Solid State and Structural Chemistry Unit,
Indian Institute of Science,
Bangalore, India 560 012.}
\end{center}

\begin{center}
{\large \bf Abstract}
\end{center}
\large

 Rotational friction on proteins and macromolecules is known to derive 
contributions from at least two distinct sources -- hydrodynamic (due to 
viscosity) and dielectric friction (due to polar interactions). 
In the existing theoretical approaches, the effect of the latter is taken 
into account in an {\it ad hoc} manner, by increasing the size of the 
protein with the addition of a hydration layer. Here we calculate the 
rotational dielectric friction on a protein ($\zeta_{DF}$) by using a 
generalized arbitrary charge distribution model (where the charges are 
obtained from quantum chemical calculation) and the hydrodynamic friction 
with stick boundary condition, ($\zeta_{hyd}^{stick}$) by using the 
sophisticated theoretical technique known as tri-axial ellipsoidal method, 
formulated by Harding [S. E. Harding, Comp. Biol. Med. {\bf 12}, 75 (1982)]. 
The calculation of hydrodynamic friction is done with only the dry volume 
of the protein (no hydration layer). We find that the total friction 
obtained by summing up $\zeta_{DF}$ and $\zeta_{hyd}^{stick}$ gives reasonable agreement 
with the experimental results, i.e., $\zeta_{exp} \approx \zeta_{DF} + 
\zeta_{hyd}^{stick}$. 

\newpage
\section{Introduction}

  In this article, we present an interesting result that the experimentally 
observed rotational correlation time of a large number of proteins can essentially 
be described as the combined effect of the rotational dielectric and hydrodynamic 
frictions on the proteins. Thus, one needs not assume the existence of a rigid 
hydration layer around the protein, as is often assumed in the standard theoretical 
calculations of hydrodynamic friction.

  The study of rotational friction of proteins in aqueous solution has a long history $[1-12]$. 
Despite many decades of study, 
several aspects of the problem remain ill understood. 
For proteins and macromolecules, the rotational friction is obtained from 
Debye-Stokes-Einstein (DSE) relation given by,
\begin{equation}
\zeta_{R} = 8\pi\eta\,R^{3}, 
\end{equation}
\noindent where $\zeta_{R}$ is the rotational friction on the protein and $R$ is 
the radius of the protein. Naturally, the above relation assumes a spherical 
shape of the protein, which is often not correct. Moreover, there is ambiguity 
about the determination of some average radius of the protein. If one obtains the 
radius from the standard mass density of the protein (0.73 gm/cc), the values of 
the rotational friction are much smaller. The dielectric measurement of 
Grant \cite{grant} showed that the experimental value of rotational friction 
of myoglobin could only be explained by the above DSE equation, if one assumes 
a thick hydration layer around the protein, thereby increasing the radius of 
the protein. It is well known that spherical approximation embedded in DSE is 
grossly in error and the shape of the protein is quite important. However, even 
with the more recent sophisticated techniques such as 
tri-axial ellipsoid method \cite{hdtriax} and the microscopic bead modeling technique 
\cite{garciabiophys,hydropro}, which take due recognition of the non-spherical shape of 
the macromolecule, agreement with the experimental result is not possible without
the incorporation of a rigid hydration layer \cite{carrasco}. In should be recognized that 
the effect of hydration layer thus introduced is purely
{\it ad hoc}. In the case of tri-axial ellipsoidal method, the values of the axes are 
increased proportionately by increasing the percentage of encapsulation of the 
protein atoms inside its equivalent ellipsoid \cite{hdhydration,muller}. On the 
other hand, the microscopic bead modeling technique uses beads of much 
bigger size \cite{garciabiophys} (3.0 \AA \, instead of 1.2 \AA) to take care of 
the effect of hydration layer. Without the hydration layer, the estimate of 
friction obtained from the theory is systematically lower.

 It has been recognized quite early that water in the hydration layer 
surrounding proteins and macromolecules has completely different dynamical 
properties than those in the 
bulk \cite{pal}. The dynamics of water molecules 
in the hydration layer are also subject of great 
interest as they could play crucial role in the property and activity of these 
molecules. One often discusses the crossover from biological activity to the 
observed inactivity at low temperatures in terms of a protein-glass transition 
observed in the hydrated proteins \cite{prtglass}. Recent investigations have shown that 
the water molecules in the hydration layer are not only more structured but they also 
show slow translational and rotational motion than their bulk 
counterpart \cite{zewail,cannistraro,marchi,steinhauser,bala}. 

 Nevertheless, it is highly unlikely that water molecules in the surface of 
a protein such as myoglobin are so slow that we can replace it by a rigid 
hydration layer. On the contrary, all the recent experimental and simulation 
studies have shown that the water in the surface of the protein exhibits 
bimodal dynamics \cite{boresch}. Majority of the water molecules seem to retain 
their bulk-like dynamics while a fraction ($\sim 20\%$) exhibits markedly 
slow dynamics. Recent solvation dynamics and photon echo peak shift 
experiment not only established the existence of slow water on the surface 
of proteins but also showed that the hydration layer is quite labile \cite{photonecho}. 
If one defines an average residence time to characterize the dynamics of water 
in the hydration layer, the residence time of bound or quasi-bound water 
is expected to range from 20 to 300 ps \cite{residence}. Question naturally 
arises how to understand quantitatively the role of the hydration layer 
in enhancing the rotational friction on the protein molecules. Clearly, 
the picturesque description of an immobile rigid layer around protein needs 
to be replaced by a description where the hydration layer is slow but 
definitely dynamic.

 This labile hydration layer has been explained in terms of a dynamic exchange 
model, which assumes that due to the presence of relatively stronger hydrogen 
bonding of water molecules with the charged groups at the surface of the protein, 
a surface water molecule can exist in either of the following two states -- 
bound and free \cite{nandibagchi}. The free water molecules have dynamical 
characteristics similar to those of the bulk but the bound water molecules are 
essentially made static by the hydrogen bonding with the surface. In this 
picture, the slow time scale arises due to the dynamical exchange between 
the two states of the water molecules. Recent computer simulations seem to 
have confirmed the essential aspects of the dynamic exchange model 
(DEM) \cite{bagroyal}.

 While the above model can provide a simple explanation of the origin of 
the observed slow dynamics, its correlation with dynamical properties of 
protein has not yet been established. This is a 
non-trivial problem as discussed below.

 The mode coupling theory (MCT) is another viable quantitative 
theory, which has been quite successful in describing translational and rational 
motion of small molecules \cite{montgomery}. This approach has also been extended to treat 
dynamics of polymer and biomolecules \cite{wolynesmct}. Let us recall a few 
of the lessons learned from the MCT of rotational friction of small molecules, 
and translational friction of ions in dipolar liquids. In both the cases, 
intermolecular dipole-dipole/ion-dipole correlations were found to play important 
role. It was also found that if one neglects the translational mode of the solvent molecules, 
then the friction on polar solute increases by several factors. It should be noted here 
that the continuum models/hydrodynamic description of rotational friction always 
ignored this translational component. In fact, this translational component 
plays a hidden role in reducing the effect of the role of molecular level solute 
solvent and solvent-solvent pair (both isotropic and orientational) correlations that
increase the value of the friction over the continuum model prediction. Thus, the issue
is rather involved. In fact, the continuum model is found to give accurate results 
due to cancellation 
of two errors: neglect of short-range correlations and neglect of translational 
contribution. In view of the above, it is thus important to note that the slow water 
molecules in the hydration layer can enhance the friction considerably. Thus, 
the classical picture of rigid, static hydration layer needs to be replaced 
by dynamic layer where the translational motion of the water molecules should 
be related to the residence time. However, only preliminary progress has been made 
in this direction. Thus, continuum models remain the only theoretical method to 
treat dielectric friction on complex molecules.

 An important issue in the calculation of the rotational friction is that
 proteins are characterized by complex charge distribution. The earliest models to 
estimate the enhanced friction on a probe, due to the interactions of its polar 
groups with the surrounding water molecules in an aqueous solution, employed a point 
dipole approximation \cite{nee,hubbardwoly,hynes}. In the simplest version of the model, the probe 
molecule is replaced by a sphere with a point dipole at the {\it center} of the 
sphere. Such an approach is reasonable for small molecules, although continuum 
model itself may have certain limitations. The situation is quite different for 
large molecules like 
proteins because the charge here is distributed over a large volume and the surface 
charges are close to the water molecules. Thus, the point dipole approximation becomes 
inapplicable to such systems.
This limitation of the early continuum models was removed by Alavi and 
Waldeck \cite{waldeck} who obtained an elegant expression for the dielectric friction 
on a molecule with extended arbitrary charge distribution. By studying several 
well-known dye molecules, they demonstrated that the extended charge 
distribution indeed has a strong effect on the dielectric friction on the 
probe molecules. The work of Alavi and Waldeck \cite{waldeck} constitutes an 
important advance in the study of dielectric friction. The role of dielectric friction 
has been studied for the organic molecules by other authors \cite{duttdiel}.

 The objective of the present work is to attempt to replace the rigid hydration layer
used in hydrodynamic calculation. To this goal, we calculate the hydrodynamic 
friction using the tri-axial method \cite{hdtriax}, in which the shape of 
a protein is mapped to 
an ellipsoid of three unequal axes -- closely representing the shape and size of 
the protein. No hydration layer is added in the calculation. We then calculate the 
dielectric friction using Alavi and Waldeck's model of generalized charge 
distribution for a large number of proteins. The friction contributions obtained 
from the above two methods are combined to obtain the total rotational friction. 
When compared, the total friction has been found to agree closely with the 
experimental result.

 We have also extended the work of Alavi and Waldeck to include multiple shells of 
water with different dielectric constants around a protein. The multiple shell 
model is introduced in concern with the experimental observation of varying 
dielectric constants of water from the hydration layer surrounding a protein to 
the bulk water. These shells have distinct dielectric properties -- both 
static and dynamic. The resulting analytical expressions can be used to obtain 
quantitative prediction of the effects of a slow layer of water molecules on 
the dielectric friction on proteins. However, the multiple shell model in the 
continuum fails since it adds up the friction in every layer. This has been 
discussed in the appendix.

\section{Results and Discussion}
  Below, we discuss the results obtained from the different aspects of rotational friction 
of proteins. The coordinates of the proteins are obtained from protein data bank (PDB) 
\cite{pdb}.

\subsection{Dielectric Friction}

 Dielectric friction is an important part of rotational friction for polar or 
charged molecules in polar solvent, because of the polarization of the solvent 
medium. The solvent molecules, being polarized by the probe, create a reaction 
filed, which opposes the rotation of the probe. 

 Many of the amino acid residues, which constitute the protein, are polar or 
hydrophilic. Therefore, in the aqueous solution, a protein and other polar molecules 
experience significant dielectric friction. There exist several theories 
\cite{nee,hubbardwoly,hynes,felderhof,wolyanrev,madden}, which account for the 
dielectric contribution to the friction. Some of these theories are continuum 
model calculation of a point charge or point dipole rotating within the spherical 
cavity. Nee and Zwanzig \cite{nee} provide an estimate of dielectric friction 
on a point dipole in terms of the dipole moment of the point dipole, dielectric 
constant of the solvent, Debye relaxation time, and the chosen cavity radius. 
Later, Alavi and Waldeck \cite{waldeck} extended this theory to incorporate the 
arbitrary multiple charge distribution of the probe molecule.

 The dielectric friction on the proteins has been calculated 
from the expression of Alavi and Waldeck for arbitrary multiple charge distribution 
model given below \cite{waldeck},
\begin{eqnarray}
\zeta_{DF} &=& {8\over R_{c}} { {\epsilon_{s}- 1}\over{(2\epsilon_{1} + 1)^2}}\,\tau_{D}
\sum_{j=1}^{N}\sum_{i=1}^{N}\sum_{l=1}^{\infty}\sum_{m=1}^{l} 
\biggl({{2l+1}\over{l+1}}\biggr){{(l-m)!}\over{(l+m)!}}
q_{i}q_{j} \biggl({{r_{i}}\over{R_c}}\biggr)^{l}\biggl({{r_{j}}\over{R_c}}\biggr)^{l}
\times \nonumber\\
& &m^{2} \,P_{l}^{m}(cos\,\theta_{i})\,P_{l}^{m}(cos\,\theta_{j})cos(m \phi_{ji})
\end{eqnarray}
\noindent where $R_c$ is the cavity radius, ($r_{i},\theta_{i},\phi_{i}$) is the 
position vector and $q_{i}$ is the partial charge of the $i$th atom. 
$P_{l}^{m}(cos(\theta_{i})$ 
is the Legendre polynomial. The maximum value of $l$ used in the Legendre 
polynomial is 50. $\epsilon_{s}$ is the static dielectric constant of the solvent. Since 
the solvent here is water, $\epsilon_{s}$ is taken to be 78 and the Debye 
relaxation time $\tau_{D}$ is taken as 8.3 picosecond (ps). 

 The partial charges ($q_{i}$) of the atoms constituting the proteins have been 
calculated using the extended Huckel model of the semi empirical calculation package 
of Hyperchem software. The dielectric friction is calculated on each of the atoms in a 
protein. The rotational frictions around X, Y and Z direction are calculated by changing the 
labels of the atom coordinates. The average dielectric constant $\zeta_{DF}^{av}$ is 
the harmonic mean of the dielectric frictions along X, Y and Z direction. Here X, Y, and Z 
denote the space fixed Cartesian coordinate of the proteins, as obtained from PDB \cite{pdb}.

 Table $1$ shows the values of dielectric friction along X, Y, Z direction and their average. 
 Continuum calculation method of the dielectric friction formulated by Alavi and Waldeck is 
dependent on the cavity radius and has been discussed in detail by them \cite{waldeck}. They 
calculated the cavity radius from the observed orientational relaxation time of the organic 
molecules. The ratios of the longest bond vector of the organic molecules to the cavity radius 
ranged from 0.75 to 0.85. In Table $1$, the calculation s of dielectric friction are 
performed using the cavity radius such that the ratio of the longest bond vector to the 
cavity radius is 0.75.

 In Table $2$, we compared the average dielectric friction for the two above 
ratios -- 0.75 
(denoted as $\zeta_{DF}^{0.75}$) and 0.85 (denoted as $\zeta_{DF}^{0.85}$). 
$\zeta_{DF}^{0.85}$ is always larger than $\zeta_{DF}^{0.75}$ since the shorter 
cavity radius will put the charges close to the surface of the cavity, thereby 
increasing the polarization of the 
solvent and hence the rotational friction of the molecule.
 
\subsection{Hydrodynamic Friction}
  The hydrodynamic rotational friction of the protein depends on its shape and size. 
Hydrodynamic friction was estimated earlier by the well-known DSE relation (Eq. 1).  
Perrin in 1936 \cite{perrin} extended the DSE theory to calculate the hydrodynamic 
friction for molecules with prolate and oblate like shapes. Both prolate and 
oblate have two unequal axes. Harding \cite{hdtriax} further extended the 
theory to calculate the hydrodynamic friction using a tri-axial ellipsoid. 
All the above theories employ stick binary condition to obtain the 
hydrodynamic friction. 

  Tri-axial ellipsoidal technique requires the construction of an equivalent ellipsoid of the 
protein. We have followed the method of Taylor {\it et al.} to construct an equivalent 
ellipsoid from the moment matrix \cite{taylor}. The eigenvalues of this equivalent ellipsoid 
are proportional to the square of the axes. So this method provides with the two axial 
ratios. We then obtained the values of the axes using the formula given by Mittelbach 
\cite{mittel}

\begin{equation}
R_{\gamma}^2 = {1\over 5} ({A^2 + B^2 + C^3})
\end{equation}
\noindent $R_{\gamma}$ is the radius of gyration and $A$, $B$ and $C$ are the three unequal 
axes of a particular protein. 

  Once the protein is represented as an ellipsoid with three principle axes, the hydrodynamic 
friction is calculated using Harding's method \cite{hdtriax,harding}. The hydrodynamic 
rotational friction of the ellipsoidal axes A, B and C are denoted as $\zeta_A$, $\zeta_B$ 
and $\zeta_C$. The above rotational friction is obtained from the series of 
equations given below \cite{harding},

\begin{eqnarray}
&&\zeta_{0} = 8 \pi \eta {ABC}, \nonumber \\
&&\zeta_A = \zeta_{0} {{2(B^2 + C^2)}\over{3ABC(B^2\alpha_{2}+C^2\alpha_{3})}},\nonumber \\
&&\zeta_B = \zeta_{0} {{2(A^2 + C^2)}\over{3ABC(A^2\alpha_{1}+C^2\alpha_{3})}},\nonumber \\
&&\zeta_C = \zeta_{0} {{2(A^2 + B^2)}\over{3ABC(A^2\alpha_{1}+B^2\alpha_{2})}}, \nonumber \\
&&\alpha_{1} = \int_{0}^{\infty} {{d\lambda}\over {(A^2 + \lambda)\Delta}}, \nonumber \\
&&\alpha_{2} = \int_{0}^{\infty} {{d\lambda}\over {(B^2 + \lambda)\Delta}}, \nonumber \\
&&\alpha_{1} = \int_{0}^{\infty} {{d\lambda}\over {(C^2 + \lambda)\Delta}}, \nonumber \\
&&\Delta = \biggl [ (A^2 + \lambda)(B^2 + \lambda)(C^2 + \lambda) \biggr ]^{1\over 2}
\end{eqnarray}
\noindent $\eta$ is the viscosity of the solvent.
 We have calculated the average of tri-axial hydrodynamic friction by taking a harmonic mean 
of the friction along three different axes, as given below,
\begin{equation}
\frac{1}{\zeta_{TR}^{av}} = {1\over 3} \biggl[ \frac{1}{\zeta_{TR}^{A}} + 
\frac{1}{\zeta_{TR}^{B}} + \frac{1}{\zeta_{TR}^{C}}\biggr]
\end{equation}

 The values of hydrodynamic friction, along three principle axes ($A$, $B$ and $C$) of the 
ellipsoid and their mean, are tabulated in the Table $3$. The $A$, $B$ and $C$ axes are 
not the same as the space fixed X, Y, Z Cartesian reference frame. 
Note that the values obtained 
from the tri-axial method are much lower than the experimental 
values. Here, we can talk about an important aspect of standard hydrodynamic approach -- 
hydration layer. One finds that hydrodynamic values of rotational friction underestimate the 
rotational friction unless the effect of hydration layer is taken into account. However, the 
effect of hydration layer is usually incorporated in an {\it ad hoc} manner, by increasing 
the percentage of encapsulation of the atoms inside the ellipsoid \cite{muller,hdhydration}. 
In this method, once the two axial ratios are obtained from the equivalent ellipsoid, the actual 
values of the axes are obtained by increasing the encapsulation of the protein atoms inside the 
ellipsoid. In the calculation presented here, the axes are obtained by equating with the radius 
of gyration. Therefore, we considered {\it no hydration layer} in this calculation of 
hydrodynamic friction. Later, we will show that this effect of hydration layer comes from the 
dielectric friction.

\subsection{Total rotational friction: Comparison with experimental results}

  

  We define the total rotational friction as the sum of dielectric friction ($\zeta_{DF}^{av}$) 
and the hydrodynamic friction without the hydration layer (i.e. tri-axial friction, 
$\zeta_{TR}^{av}$) as given below,
\begin{equation}
\zeta_{total} = \zeta_{DF}^{av} + \zeta_{TR}^{av}
\end{equation}

 In Table $4$, we have shown the values of the average dielectric ($\zeta_{DF}^{av}$), 
hydrodynamic ($\zeta_{TR}^{av}$) friction. Total friction ($\zeta_{total}$) defined 
above is shown in the fourth column. To compare with the experimental results, 
we have shown the experimental values of the rotational friction in the next 
column.  Note here, while the total friction, which is the contribution from both 
dielectric and hydrodynamic friction, is close to the experimental result, the 
microscopic bead modeling predicts the result, which is close to experimental 
value by itself \cite{hydropro}. The last column of Table $4$ shows the references 
of the articles from which the experimental results are obtained. 

 The similarity between the total friction and the experimental friction is shown in figure 
 $1$, where we have plotted the experimental values of rotational friction against the 
total friction for a large number of proteins. For most of the proteins, 
the results fall on the diagonal line.  

 From the results shown in Table $4$, we can conclude that the sum of dielectric friction 
and the hydrodynamic friction of the dry protein is approximately equal to the 
experimental results.

\begin{equation}
\zeta_{total} \approx \zeta_{exp}
\end{equation}

\section{Conclusion}

  Let us first summarize the main results of this work. 
We have calculated the hydrodynamic rotational friction on proteins using 
the tri-axial ellipsoid method, formulated by Harding \cite{hdtriax}, and the 
dielectric friction 
using the generalized charge distribution model derived by Alavi and 
Waldeck \cite{waldeck}. The 
hydrodynamic friction is calculated without the inclusion of any hydration layer. We 
have found that the combined effect of dielectric and hydrodynamic friction gives an estimate 
close to the experimental result. This approach seems to provide a microscopic basis for the 
standard hydrodynamic approach, where a hydration layer is added to the protein in an 
{\it ad hoc} manner, to calculate rotational friction. 

 The calculations adopted here are still not without limitations. The continuum 
calculation of dielectric friction is dependent on the assumed cavity radius. 
Unfortunately, there is yet no microscopic basis 
to assume certain value of the cavity radius for the calculation of dielectric friction. 
Moreover, the effect of increasing dielectric constant of the solvent from the 
vicinity of the protein to the bulk is not taken into account by Alavi and 
Waldeck \cite{waldeck}. Thus, we have attempted to incorporate a multi shell model to 
incorporate multiple shells with varying dielectric constants. The 
theory is described in the appendix in detail. The drawback of incorporation 
of multiple shells in the continuum is that the frictional contributions from 
each of the shells add up, thereby giving rise to an unphysical large 
result.

 Similarly, the tri-axial method and bead modeling method suffer from the lack of 
microscopic basis to determine the exact values of the axes and the bead size, respectively.

 A potentially powerful approach to the problem is the mode coupling theory 
\cite{bcacp}, which uses the time correlation formalism to obtain the memory kernel 
of the rotational friction. 
The total torque is separated into two parts -- a short range part (which is called the 
bare friction $\Gamma_{bare}$) and a long range dipolar part. The advantage of mode coupling 
theory is that it does not depend on any parameter. It uses a time dependent 
effective potential field in terms of density distribution and the direct correlation 
function given by \cite{bcacp},
\begin{equation}
V_{eff}({\bf r},{\Omega},t) = -k_{B}T \int dr^{'}d \Omega^{'} c({\bf r} - 
{\bf r^{'}},\Omega,\Omega^{'})\nabla_{\Omega} \rho(r^{'},\Omega^{'},t)
\end{equation}
The torque density is then expressed as,
\begin{equation}
{\bf N}_{c}({\bf r},\Omega,t) = n({\bf r},\Omega,t) \biggl [ 
-\nabla_{\Omega}V_{eff} ({\bf r},{\Omega},t) \biggr ]
\end{equation}
\noindent where, $n({\bf r},\Omega,t)$ is the number density of the tagged particle. 
The rotational friction comes from the torque-torque correlation function. The final 
expression of the single particle ($\Gamma_{s}$) and collective friction ($\Gamma_{c}$) are 
given by \cite{bagchimol},

\begin{equation}
\Gamma_{s}(z) = \Gamma_{bare} + {\cal A} \int_{0}^{\infty} e^{-zt} \int_{0}^{\infty} 
dk \: k^{2} \sum_{l_{1}l_{2}m} c_{l_{1}l_{2}m}^{2}(k)F_{l_{2}m}(k,t)
\end{equation}

\begin{equation}
\Gamma_{c}(z) = \Gamma_{bare} + {\cal A} \int_{0}^{\infty} e^{-zt} \int_{0}^{\infty} 
dk \: k^{2} \sum_{l_{1}l_{2}m} F_{l_{1}m}^{s}(k,t) \,\: 
c_{l_{1}l_{2}m}^{2}(k) \,\: F_{l_{2}m}(k,t)
\end{equation}
\noindent where, ${\cal A} = {\rho \over {2 \,(2\pi)^{4}}} $ . $c_{l_{1}l_{2}m}$ 
is the $l_{1}l_{2}m$-th coefficient of the two particle direct correlation 
function between any two dipolar molecules. 
$F_{l_{1}l_{2}m}^{s}$ and $F_{l_{2}m}(k,t)$ are the single particle and the 
collective orientational correlation functions, respectively.

 Eq. 10 and Eq. 11 are the standard mode coupling theory expressions for rotational friction. 
It has to be solved self consistently. In the overdamped limit, the self 
dynamic structure factor is expressed as,
\begin{equation}
F_{lm}^{s}(k,z) =  \biggl[z + \frac{k_{B}T l(l+1)}{I \Gamma_{s}(z)}\biggr ]^{-1}
\end{equation}
and the collective dynamic structure factor is given by,
\begin{equation}
F_{lm}^{c} = F_{lm}(k) \biggl[z + \frac{k_{B}T f_{llm}(k)l(l+1)}{I \Gamma_{c}(z)} 
+ \frac{k_{B}T k^{2}f_{llm}(k)}{M \Gamma_{T}(z)}\biggr ]^{-1}
\end{equation}
\noindent where $f_{llm}(k) = 1 - (-1)^{m}(\rho/4\pi)c_{llm}(k)$. $I$ and $M$ are the 
moment of inertia and the mass of the dipolar molecule, respectively. $\Gamma_{T}(z)$ is 
the frequency dependent translational friction. 

 The advantage of the mode coupling approach is that the once the charge density of the 
protein molecules and the dipole density of the water molecules surrounding the protein 
are defined, the rotational friction can be obtained in terms of the direct correlation 
function and the static and dynamic structure factors of the protein-water systems. These 
are again related by Ornstein-Zernike 
equation \cite{gray}. 

 The important aspect of this microscopic theory of dielectric friction is the hidden 
contribution of the translational modes. In the hydration layer, the rotational friction 
is enhanced due to the slow translational component. This effect of translation 
could not be approached through continuum calculation. Work in this direction is under 
progress. 

\section{Appendix : Multiple shell model and the Drawback}
 
 Dielectric constant of water varies from the vicinity of the protein to the 
bulk water value. To understand the effect of this varying dielectric constant 
on the rotational dielectric 
friction of the protein, we have performed the continuum calculation of rotational 
dielectric friction using a multiple shell model.

 Nee and Zwanzig derived the dielectric friction contribution of a point dipole 
\cite{nee}. Alavi {\it et al.} \cite{waldeck} generalized it to obtain dielectric 
friction of a molecule with arbitrary distribution of charges. Castner {\it et al.} 
\cite{castner} generalized the point dipole approach to incorporate the discrete 
shell model with varying dielectric constant. Here, we have combined the 
approach of Castner {\it et al.} and Alavi {\it et al.} to obtain a generalized 
arbitrary charge distribution model for multiple hydration layers with varying 
dielectric constants around the protein. 
{\bf Figure $2$} shows the general scheme of this work. The protein is in the 
innermost cavity of radius $a$, where the water has a dielectric constant value 
of 4. The dielectric constant of water in the successive layers is assumed to 
increase up to the value of the bulk water, having a dielectric constant of 78.
The width of each shell is assumed to be $d$.

 We first write down the electrostatic potential in two dimensions, which could 
be generalized to three dimensions using principle of superposition. 
The electrostatic potential $\Phi_{j}({\bf r})$ can be written as,
\begin{equation}
\Phi_{j}({\bf r}) = \Phi_{j}(r,\theta) = \sum_{l=0}^{\infty} B_{l}^{j}\,\,
{{P_{l}(cos \, \theta)}\over{r^{l+1}}} \,\, +\,\, A_{l}^{j} \,\, r^{l} \, P_{l}(cos\,\theta) 
\end{equation}
\noindent where, $j$ denotes the number of concentric shells surrounding the protein. 
For $n$ concentric shell, $j$ can have a value from $0$ to $n+1$. 
$j=0$ denotes no boundary. 
The boundary conditions are,\\
\noindent(i)  $B_{l}^{0} = {q_{i}r_{i}^{l}}$, where $q_{i}$ and $r_{i}$ are 
the partial charge and the position of the $i$th atom, respectively.\\ 
(ii) $\Phi \rightarrow 0 \;\;\;\;$ as $ r \rightarrow \infty $ \\
(iii)$\Phi_{j}(r_{j}) = \Phi_{j+1}(r_{j+1})$, for $j=0,1,2...n$.\\
(iv) $\epsilon_{j}\Phi_{j}^{'}(r_{j}) = \epsilon_{j+1}\Phi_{j+1}^{'}(r_{j})$, 
for $j=0,1,2..n$, $\Phi_{j}^{'}(r) = {\frac{\partial \Phi_{j}(r)}{\partial r}}$. \\
(v ) $A_{l}^{n+1} = 0$,\\
After incorporating the boundary conditions in Eq. 14, we get,
\begin{equation}
A_{l}^{j} = -\sum_{k=j}^{n} {{B_{l}^{k}}\over{r_{k}^{2l+1}}} \times \biggl ( 
{ {\epsilon_{k+1}/\epsilon_{k} - 1} \over {\epsilon_{k+1}/\epsilon_{k} + {l\over{l+1}}}}\biggr )
\end{equation}
\noindent where, 
\begin{equation}
B_{l}^{j} = q_{i}\; r_{i}^{l}\; \biggl ( {{2l+1}\over{l+1}}\biggr )^{j} \Pi_{k=1}^{j} \biggl ( 
{1\over{\epsilon_{k}/\epsilon_{k-1}} + {l\over{l+1}}}\biggr )
\end{equation}

The reaction potential is given by,\\
\begin{equation}
\Phi_{j}(r,\theta,\phi) = \sum_{i=1}^{N}\sum_{l=0}^{\infty} A_{l}^{0}\; r^{l} 
\; P_{l}(cos\,\gamma_{i})
\end{equation}
\noindent where,
\begin{equation}
P_{l}(cos\,\gamma_{i}) = {{4\pi}\over{2l+1}} \sum_{m=-l}^{m=l} Y_{l}^{m*}(\theta_{i},\phi_{i}) 
Y_{l}^{m}(\theta_{i},\phi_{i})
\end{equation}
After few steps of algebra, we obtain the frequency dependent dielectric friction given below,
\begin{eqnarray}
\zeta_{l}^{m}(\omega) &=& \sum_{j=1}^{N}\sum_{i=1}^{N}\sum_{l=1}^{\infty}\sum_{m=1}^{l} 
{2{q_{i}q_{j}}\over{a\,\omega}}\biggl({{r_{i}}
\over{a}}\biggr)^{l}\biggl({{r_{j}}\over{a}}\biggr)^{l}\times \nonumber\\
& & {\frac{(l-m)!}{(l+m)!}}P_{l}^{m}(cos \theta_{i})P_{l}^{m}(cos \theta_{j})\: m \: 
cos(m\phi_{ji})  \times \nonumber\\
& & \sum_{s=0}^{n} \biggl |Im \biggl [\biggl({a \over{a+sd}}\biggr)^{2l+1}\biggl({{\epsilon_{s+1,s}(m\,\omega)- 1}\over{\epsilon_{s+1,s}(m\,\omega) + {l\over {l+1}}}}\biggr)\times\nonumber\\
& &\Pi_{k=1}^{s} \biggl(1- {{\epsilon_{k,k-1}(m\,\omega)- 1}\over{\epsilon_{k,k-1}(m\,\omega) 
+ {l\over{l+1}}}} \biggr) \biggr]\biggr |
\end{eqnarray}
\noindent where, $\epsilon_{j,j-1} = \epsilon_{j}/\epsilon_{j-1}$, for all values of $j$. 
$\epsilon_{0}$ is the dielectric constant of the cavity. 

 Above is the general expression of multiple ($n$) shell model. To write the final expression of 
dielectric friction for a two shell model, we assume Debye relaxation for the frequency 
dependent dielectric friction of two shells as given below,

\begin{equation}
\epsilon_{1,0}(m\omega) = 1 + {{\epsilon_{1,0} -1}\over{1+ i\,m\omega\,\tau_{D1}}}
\end{equation}

\begin{equation}
\epsilon_{2,1}(m\omega) = 1 + {{\epsilon_{2,1} -1}\over{1+ i\,m\omega\,\tau_{D2}}}
\end{equation}
\noindent where, $\tau_{D1}$ and $\tau_{D2}$ are the Debye relaxation time for 
the first and second shell.

The expression of dielectric friction for a two-shell model is given below,

\begin{eqnarray}
\zeta_{DF} &=& \sum_{j=1}^{N}\sum_{i=1}^{N}\sum_{l=1}^{\infty}\sum_{m=1}^{l} 
{8\over a}\, \biggl({{2l+1}\over{l+1}}\biggr)\, {{(l-m)!}\over{(l+m)!}}\,\,
q_{i}q_{j} \,\,\biggl({{r_{i}}
\over{a}}\biggr)^{l}\,\,\biggl({{r_{j}}\over{a}}\biggr)^{l}\times \nonumber\\
& &m^{2} \,\, P_{l}^{m}(cos\,\theta_{i})\,\, P_{l}^{m}(cos\,\theta_{j})\,\, 
cos(m\phi_{ji})\times \nonumber\\
& &\biggl[{ {\epsilon_{1,0}- 1}\over{(2\epsilon_{1,0} + 1)^2}}\,\tau_{D1}\;\;+\;\; 
\biggl({a\over{a+d}}\biggr)^{2l+1} {{\epsilon_{2,1}-1} \over {(2\epsilon_{2,1}+ 1)^2}}
\,\tau_{D2} \nonumber\\
& &+2\,\biggl({a\over{a+d}}\biggr)^{2l+1} { {\epsilon_{1,0}- 1}\over{(2\epsilon_{1,0} + 1)^2}}
\;{ {\epsilon_{2,1}- 1}\over{(2\epsilon_{2,1} + 1)^2}}\times \nonumber \\
& &\biggl\{({2\epsilon_{1,0}+1})\tau_{D2}+ ({2\epsilon_{2,1}+1})\tau_{D1}\biggr\} \biggr], 
\end{eqnarray}
 
 The above expression has been numerically evaluated to find out the effect of dielectric friction 
on protein due to varying dielectric constant of water around the protein. The multiple shell model 
is found to overestimate the dielectric friction, as is evident from the above expression.

{\large \bf Acknowledgment}
The work is supported by DST, DBT and CSIR. A.M. thanks CSIR for Senior Research 
Fellowship.



\newpage
\begin{center}
{\bf Table 1}\\
\end{center}
{\bf Table for the dielectric friction. The unit is $10^{-23}$ erg-sec}. 
Cavity radius is chosen such a way that the ratio of longest bond vector 
($R_{max}$) of the protein to the chosen cavity radius ($R_{C}$) is 0.75. 
\begin{center}
\begin{tabular}{||c|c|c|c|c|c||}\hline
Molecule & $R_{C}$ (\AA) & $\zeta_{DF}^{X}$ & $\zeta_{DF}^{Y}$ & 
$\zeta_{DF}^{Z}$ & $\zeta_{DF}^{av}$\\\hline   
6pti  &     29.50  &      17.8  &      13.2  &      18.1  &      16.0\\\hline
1ig5  &     26.10  &      43.3  &      36.6  &      39.1  &      39.5\\\hline
1ubq  &     34.30  &      18.1  &      18.3  &      21.8  &      19.3\\\hline
351c  &     25.50  &      52.3  &      41.0  &      41.9  &      44.5\\\hline
1pcs  &     27.20  &      90.5  &      51.3  &      66.1  &      65.7\\\hline
1a1x  &     33.10  &      63.0  &      68.9  &      49.5  &      59.3\\\hline
1gou  &     32.20  &      43.8  &      67.8  &     103.6  &      63.5\\\hline
1aqp  &     35.30  &      44.5  &      71.1  &     132.1  &      68.0\\\hline
1e5y  &     33.10  &      98.9  &      70.6  &      89.9  &      84.7\\\hline
1bwi  &     35.70  &      78.3  &      60.5  &     108.1  &      77.8\\\hline
1b8e  &     33.50  &     113.3  &     112.2  &     110.5  &     112.0\\\hline
4ake  &     50.30  &      76.1  &     170.8  &     123.4  &     110.7\\\hline
3rn3  &     35.00  &     118.8  &      89.0  &      56.8  &      80.5\\\hline
1mbn  &     28.00  &     170.7  &     162.0  &     160.6  &     164.3\\\hline
\end{tabular}
\end{center}
\newpage

\begin{center}
{\bf Table 2}\\
\end{center}
{\bf Cavity size dependence of the dielectric friction. The unit is $10^{-23}$ erg-sec}.
\begin{center}
\begin{tabular}{||c|c|c|c||}\hline
Molecule & $\zeta_{DF}^{0.75}$ & $\zeta_{DF}^{0.85}$ \\\hline
6pti     &   16.4              &   25.7  \\\hline
1ig5     &   39.7              &   61.3  \\\hline
1ubq     &   19.4              &   30.3  \\\hline
351c     &   45.1              &   69.3  \\\hline
1pcs     &   69.3              &  111.0  \\\hline
1a1x     &   60.5              &   96.4  \\\hline
1gou     &   71.7              &  114.6  \\\hline
1aqp     &   82.6              &  132.3  \\\hline
1e5y     &   86.5              &  136.4  \\\hline
1bwi     &   82.3              &  128.9   \\\hline
1b8e     &  112.0              &  174.1   \\\hline         
4ake     &  123.4              &  211.7   \\\hline
3rn3     &   88.2              &  138.1   \\\hline
1mbn     &  164.5              &  263.1  \\\hline
6lyz     &  107.8              &  172.7  \\\hline
\end{tabular}
\end{center}

\newpage
\begin{center}
{\bf Table 3}\\
\end{center}
{\bf Table for the stick hydrodynamic friction using tri-axial ellipsoid. The unit is $10^{-23}$ erg-sec}.
\begin{center}
\begin{tabular}{||c|c|c|c|c|c||}\hline
Molecule & $R_{\gamma}$ (\AA) & $\zeta_{TR}^{A}$ & $\zeta_{TR}^{B}$ & 
$\zeta_{TR}^{C}$ & $\zeta_{TR}^{av}$\\\hline  
6pti  &     11.34  &      57.8  &      83.4  &      85.1  &      73.1\\\hline
1ig5  &     11.36  &      72.9  &      78.9  &      84.9  &      78.6\\\hline
1ubq  &     11.73  &      71.2  &      89.9  &      94.0  &      83.8\\\hline
351c  &     11.51  &      77.3  &      84.5  &      85.3  &      82.2\\\hline
1pcs  &     12.38  &      78.9  &     106.5  &     111.3  &      96.6\\\hline
1a1x  &     13.47  &     120.8  &     127.3  &     143.8  &     129.9\\\hline
1gou  &     13.61  &     103.3  &     141.7  &     148.2  &     127.7\\\hline
1aqp  &     14.45  &     117.7  &     171.1  &     177.0  &     150.1\\\hline
1e5y  &     13.81  &     108.9  &     145.7  &     155.3  &     133.4\\\hline
1bwi  &     13.94  &     106.9  &     155.4  &     158.2  &     135.7\\\hline
1b8e  &     14.70  &     167.5  &     172.5  &     178.2  &     172.6\\\hline
4ake  &     19.59  &     298.3  &     422.7  &     442.8  &     376.1\\\hline
3rn3  &     14.31  &     112.9  &     166.5  &     172.2  &     145.1\\\hline
1mbn  &     15.25  &     163.7  &     181.2  &     210.1  &     183.1\\\hline
\end{tabular}
\end{center}

\newpage
\begin{center}
{\bf Table 4}\\
\end{center}
{\bf Comparison between the total friction and the experimental results. Results 
are given in the unit of $10^{-23}$ erg-sec}. The references to the experimental 
results of rotational diffusion of the corresponding proteins are given in 
the Ref. \cite{halle} .
\begin{center}
\begin{tabular}{||c|c|c|c|c|c||}\hline
Protein  & PDB id & $\zeta_{DF}^{av}$ & $\zeta_{TR}^{av}$ & $\zeta_{total}$ 
& $\zeta_{exp}$ \\\hline
Bovine pancreatic trypsin inhibitor & 6pti  &      16.0  &      73.1  &      89.1	& 96.8 \\\hline
Calbindin D9k, holo form            & 1ig5  &      39.5  &      78.6  &     118.1	& 125.0 \\\hline
Human ubiquitin                     & 1ubq  &      19.3  &      83.8  &     103.1	& 118.9 \\\hline
Ferricytochrome c$_{551}$           & 351c  &      44.5  &      82.2  &     126.7	& 130.1 \\\hline
Plastocyanin, Cu(II) form           & 1pcs  &      65.7  &      96.6  &     162.3	& 149.5 \\\hline
Oncogenic protein p13$^{MTCP1}$     & 1a1x  &      59.3  &     129.9  &     189.2	& 241.9 \\\hline
Binase                              & 1gou  &      63.5  &     127.7  &     191.2	& 191.3 \\\hline
Ribonuclease A                      & 1aqp  &      68.0  &     150.1  &     218.1	& 186.1 \\\hline
Azurin, Cu(I) form                  & 1e5y  &      84.7  &     133.4  &     218.1	& 190.4 \\\hline
Hen egg-white lysozyme              & 1bwi  &      77.8  &     135.7  &     213.5	& 203.6 \\\hline
Bovine -lactoglobulin, monomer      & 1b8e  &     112.0  &     172.6  &     284.6	& 270.6 \\\hline
Adenylate kinase, apo form          & 4ake  &     110.7  &     376.1  &     486.8	& 478.2 \\\hline
Bovine Ribonuclease A               & 3rn3  &      80.5  &     145.1  &     225.6	& 235.0 \\\hline
Sperm Whale Myoglobin               & 1mbn  &     164.3  &     183.1  &     347.4	& 246.3 \\\hline
\end{tabular}
\end{center}

\begin{figure}
\centerline{\includegraphics[height=10cm]{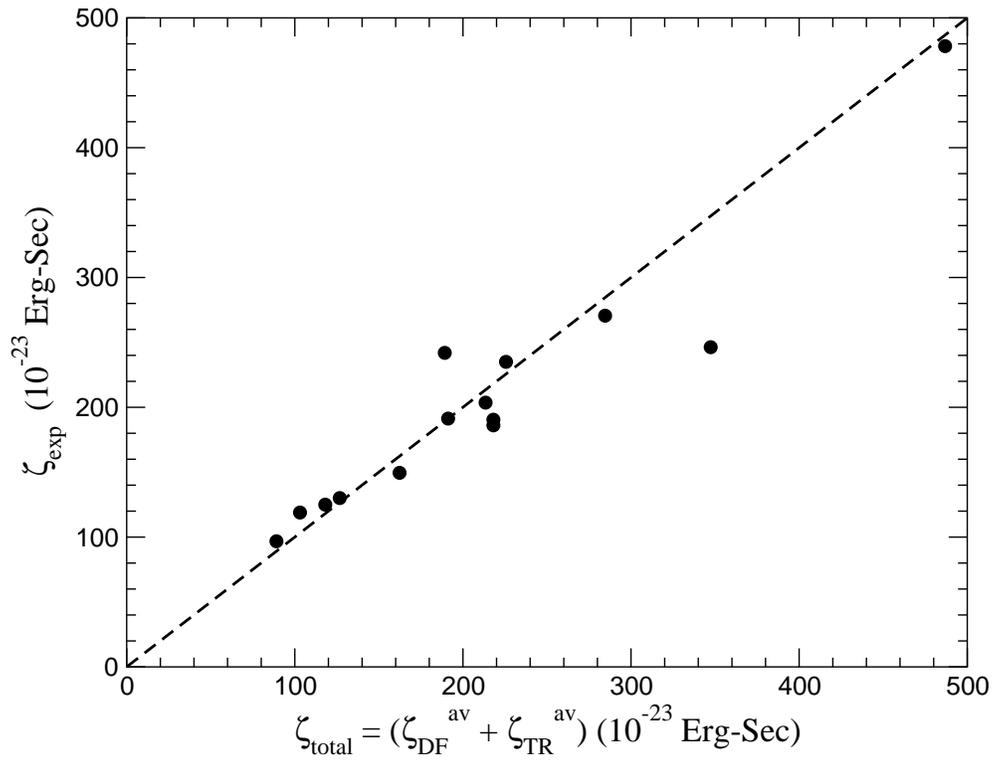}}
\caption{The combined friction from hydrodynamic and dielectric is plotted against 
the experimental results. The solid line shows the diagonal to guide the eye. }
\end{figure}

\clearpage

\begin{figure}
\centerline{\includegraphics[height=8cm]{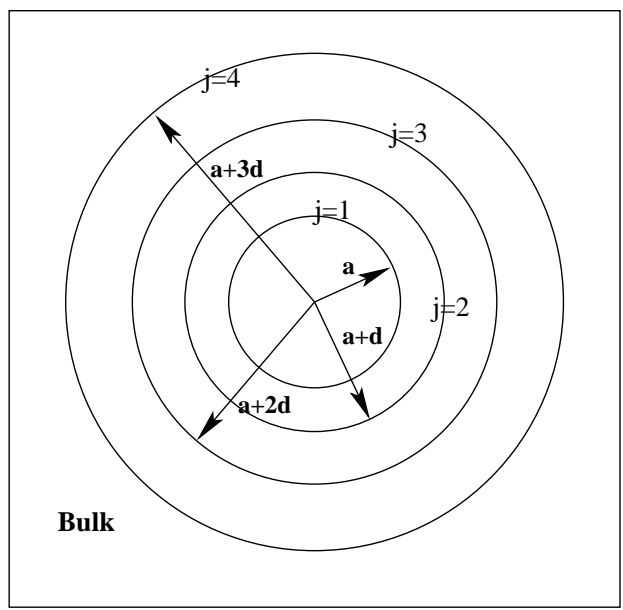}}
\caption{schematic diagram of the Molecular cavity and the hydration shell constituted by the 
bound water molecules. The bulk water molecules are more randomly oriented}
\end{figure}

\end{document}